\documentclass[11pt]{article}
\usepackage{moriond,epsfig}
\usepackage{graphicx}
\usepackage[figuresright]{rotating}
\usepackage{wrapfig}

\newcommand{\beq}{\begin{equation}}
\newcommand{\eeq}{\end{equation}}

\def\be{\begin{equation}}
\def\ee{\end{equation}}
\def\bea{\begin{eqnarray}}
\def\eea{\end{eqnarray}}

\begin{document}
\vspace*{4cm}
\title{RELATIVISTIC HEAVY ION PHYSICS: \\ A THEORETICAL OVERVIEW\footnote{Invited talk given at the XXXIXth Rencontres de Moriond 
Conference on "QCD and High Energy Hadronic Interactions", La Thuile, Italy, March 28 - April 4, 2004.} }

\author{ D. KHARZEEV }

\address{Department of Physics, Brookhaven National Laboratory,\\
Upton, New York 11973-5000, USA}

\maketitle\abstracts{
This is a mini--review of recent theoretical work in the field of relativistic 
heavy ion physics. The following topics are discussed: initial conditions and 
the Color Glass Condensate; approach to thermalization and the hydrodynamical evolution; hard probes and the properties of the Quark--Gluon Plasma. Some of the unsolved problems and potentially promising directions 
for future research are listed as well.}

\section{Introduction}

In general, theorists get attracted to relativistic heavy ion physics because 
it is placed at the intersection of three different, and equally interesting, directions in 
contemporary theoretical research: 
i) small $x$, high parton density QCD;  ii) non-equilibrium field theory; and 
iii) phase transitions in strongly interacting matter.
Indeed, understanding the evolution of a heavy ion collision requires a working theory 
of initial conditions, of the subsequent evolution of the produced partonic system, and 
 of the phase transition(s) to the deconfined phase.  
This mini--review is an attempt to capture some of the recent changes and developments in the theoretical picture of these 
phenomena which have been triggered by an intense stream of the new data from RHIC.  

\section{Initial conditions and global observables}

\subsection{The r\^{o}le of coherence}

Not so long ago, before the advent of RHIC, it was widely believed that at collider 
energies the total multiplicities will become dominated by 
hard incoherent processes.  \begin{figure}[htb]
\noindent
\vspace{-0.3cm}
\begin{minipage}[b]{.46\linewidth}
\includegraphics[width=8.3cm]{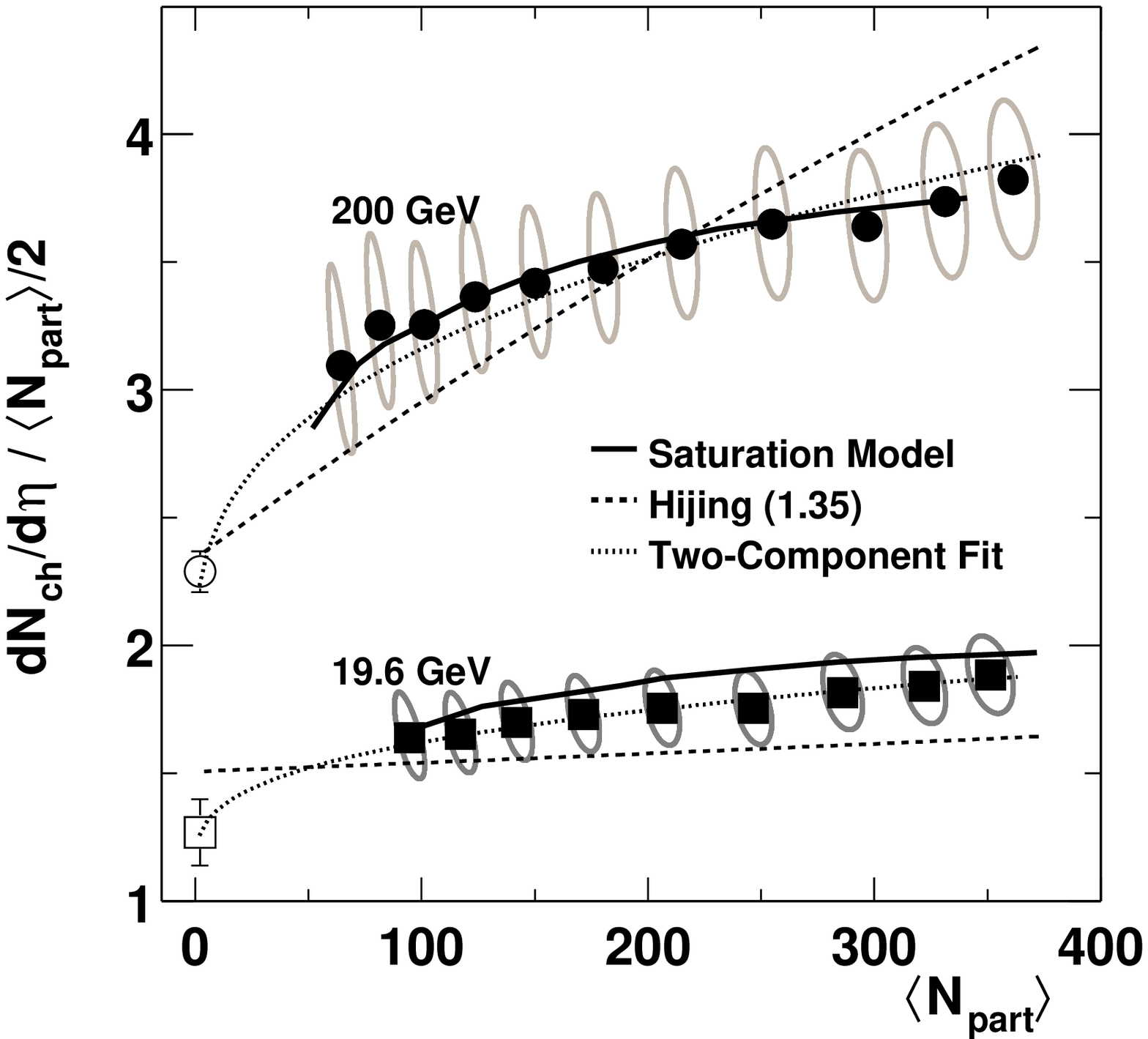}
\end{minipage}\hfill
\begin{minipage}[b]{.46\linewidth}
\includegraphics[width=8.3cm]{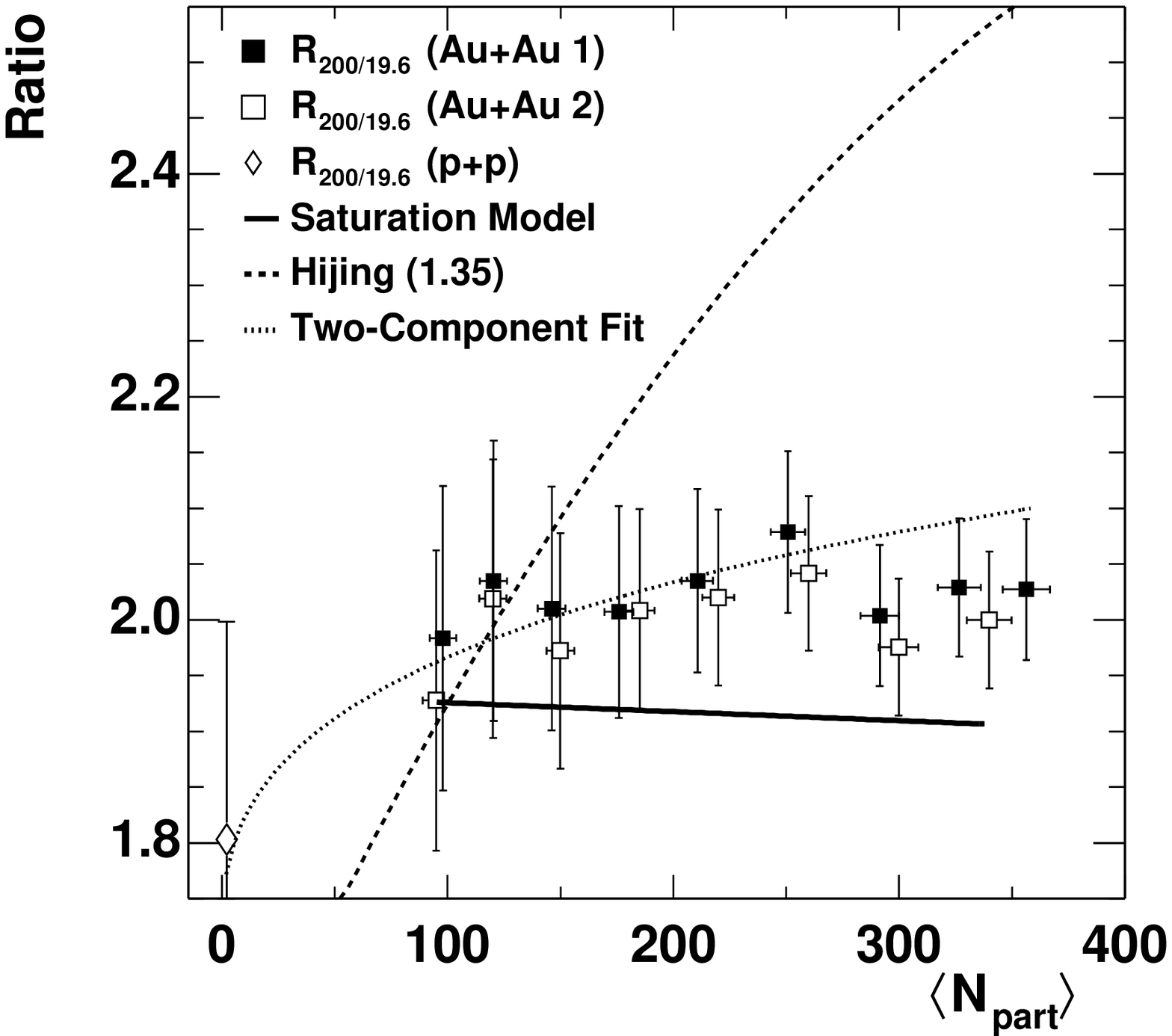}
\end{minipage}
\caption{
Centrality dependence of the charged particle multiplicity near  mid-rapidity in Au + Au 
collisions at $\sqrt{s} = 20$ and $200$ GeV; from\protect\cite{Back:2004dy}.}
\vspace{-0.5cm}
\label{fig:phob20}
\end{figure}
The very first data from RHIC (see \cite{Hofman:2004pa} and references therein in this volume) provided a lot of food for new thought: the measured charged hadron multiplicities in $Au-Au$ collisions 
appeared much smaller than expected on the basis of incoherent superposition of hard processes.
Given that any inelastic rescatterings in the final state can only 
increase the multiplicity\footnote{For statistical systems, this is due to the second law of thermodynamics}, 
we have an experimental {\it proof} of a high degree 
of coherence in multi-particle production in nuclear collisions at RHIC energies.


\subsection{Semi--classical QCD and hadron multiplicities}

Combining the idea of coherence with the parton model, we have to consider the initial parton wave functions of the 
colliding nuclei as coherent superpositions of the wave functions of the constituent nucleons. Since at small Bjorken $x$ all of the partons in the nucleus 
at a fixed transverse coordinate participate in a hard scattering process, this treatment naturally 
leads to the notion of parton density in the transverse plane $Q_s^2$ -- a new dimensionful scale of the problem. Once this scale becomes 
comparable to the resolution scale determined by the kinematics of the hard scattering, the amplitude of the process is severely affected 
by the coherence. The limit on the parton density is reached when the occupation numbers of the gluon field modes 
with transverse momenta 
$p_T < Q_s$ 
reach the value $n_k \sim 1/\alpha_s(Q_s)$, characteristic for classical gauge fields -- this is the phenomenon known as 
"parton saturation"\cite{Gribov:1984tu}, leading to a coherent state of gluons -- Color Glass Condensate (for reviews, see \cite{McLerran:2003ca,Iancu:2003xm,Mueller:2001fv,Levin:2001eq,Kharzeev:2002np}).

Since the integrated multiplicities are dominated by momenta $p_T \leq Q_s$ and parton density in the transverse plane scales 
as $Q_s^2 \sim N_{part}^{1/3}$ (where $N_{part}$ is the number of nucleons which participate in the process), Color Glass Condensate leads to a simple 
prediction\cite{Kharzeev:2000ph} for the centrality dependence of hadron multiplicity in heavy ion collisions: 
\beq 
{d n_{AA} \over d \eta} \sim N_{part}  \ \ln\left(N_{part}\right).
\eeq
Combined with the dependence of the gluon structure function on Bjorken $x$ known from HERA, which implies $Q_s^2(x) \sim 1/x^{\lambda}$, 
one can generalize this formula to predict the energy, centrality, rapidity, and atomic number dependencies of hadron 
multiplicities\cite{Kharzeev:2000ph}. Additional information on the dynamics of the collision can be inferred from the numerical lattice simulations\cite{KV}. 
So far this approach has been quite successful in predicting the multiplicities measured at RHIC; a recent important example is given at Fig.\ref{fig:phob20} which shows the evolution of centrality dependence with energy in the entire RHIC range between $\sqrt{s} = 20$ and $200$ GeV.   
One can see that the shape of the centrality dependence changes very little over a large energy range, in which 
the perturbative minijet cross section grows by over an order of magnitude. The prediction of 
the saturation model \cite{Kharzeev:2000ph} is seen to agree with the data reasonably well; this indicates the possibility 
that parton saturation sets in in heavy ion collisions already at moderate energies.  We do not expect the method to apply below $\sqrt{s} = 20$ GeV 
however, since at lower energies the coherence length becomes shorter than the nuclear radius. 

\subsection{High $p_T$ hadron suppression at forward rapidities, and quantum evolution in the Color Glass Condensate}

Parton saturation at transverse momenta $k_T \leq Q_s$ at sufficiently small $x$ appears to have non-trivial consequences also for the 
nuclear dependence of the semi-hard processes. At very small $x$, when $\alpha_s \ln 1/x \sim 1$, a semi-classical description has to be modified due to the quantum evolution.
\begin{wrapfigure}{r}{0.6 \textwidth}
\vspace{-1.3cm}
\begin{minipage}[b]{.35\linewidth}
\includegraphics[width=6cm]{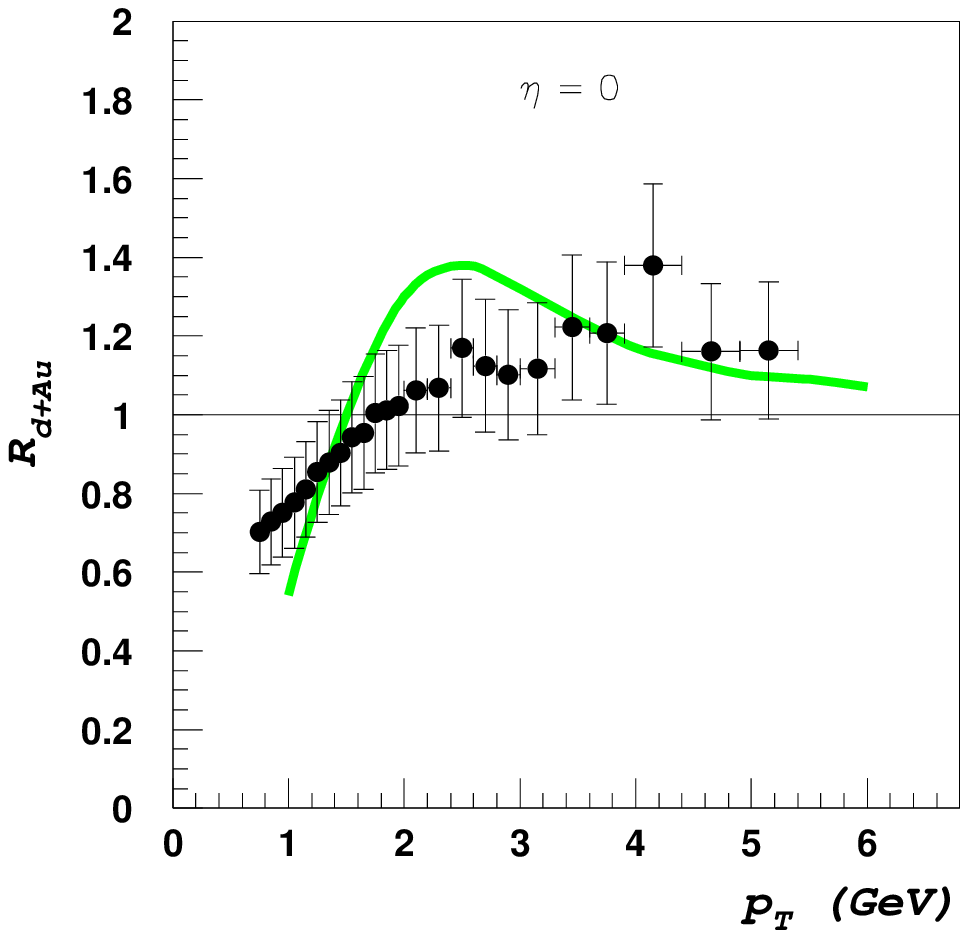}
\end{minipage}\hspace{1cm}
\begin{minipage}[b]{.35\linewidth}
\includegraphics[width=6cm]{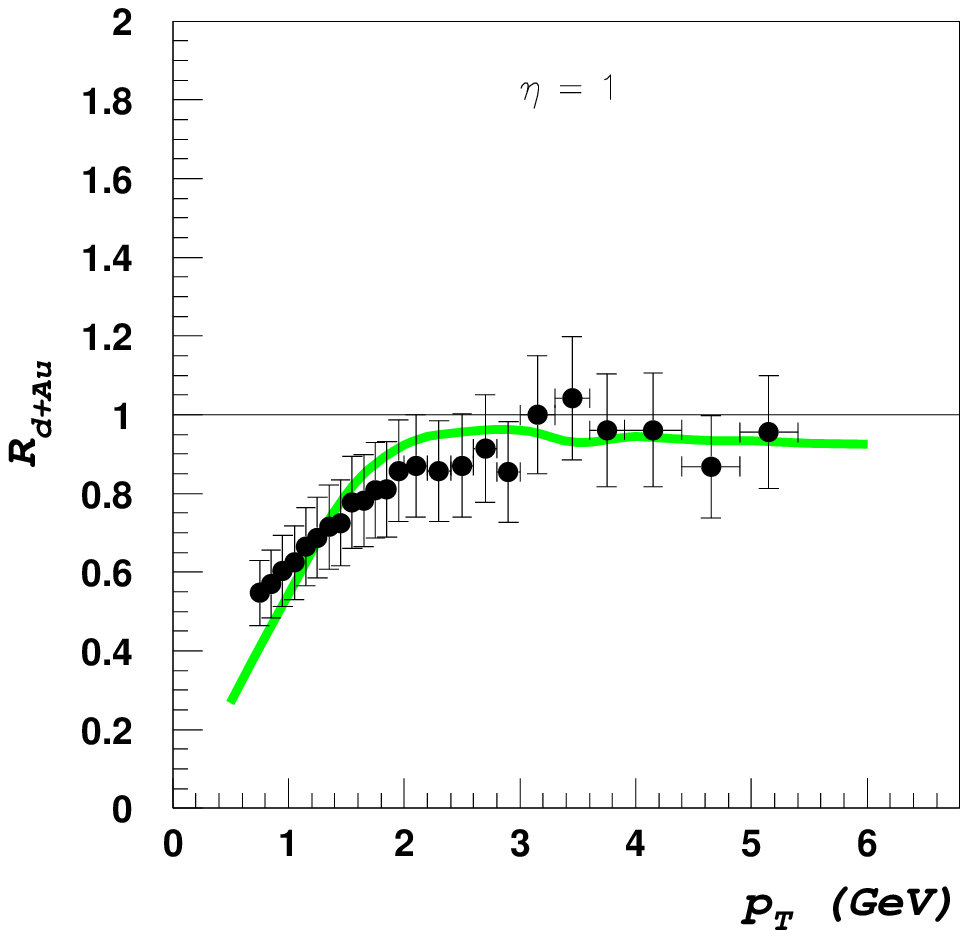}
\end{minipage}

\vspace{-0.8cm}
\begin{minipage}[b]{.35\linewidth}
\includegraphics[width=6cm]{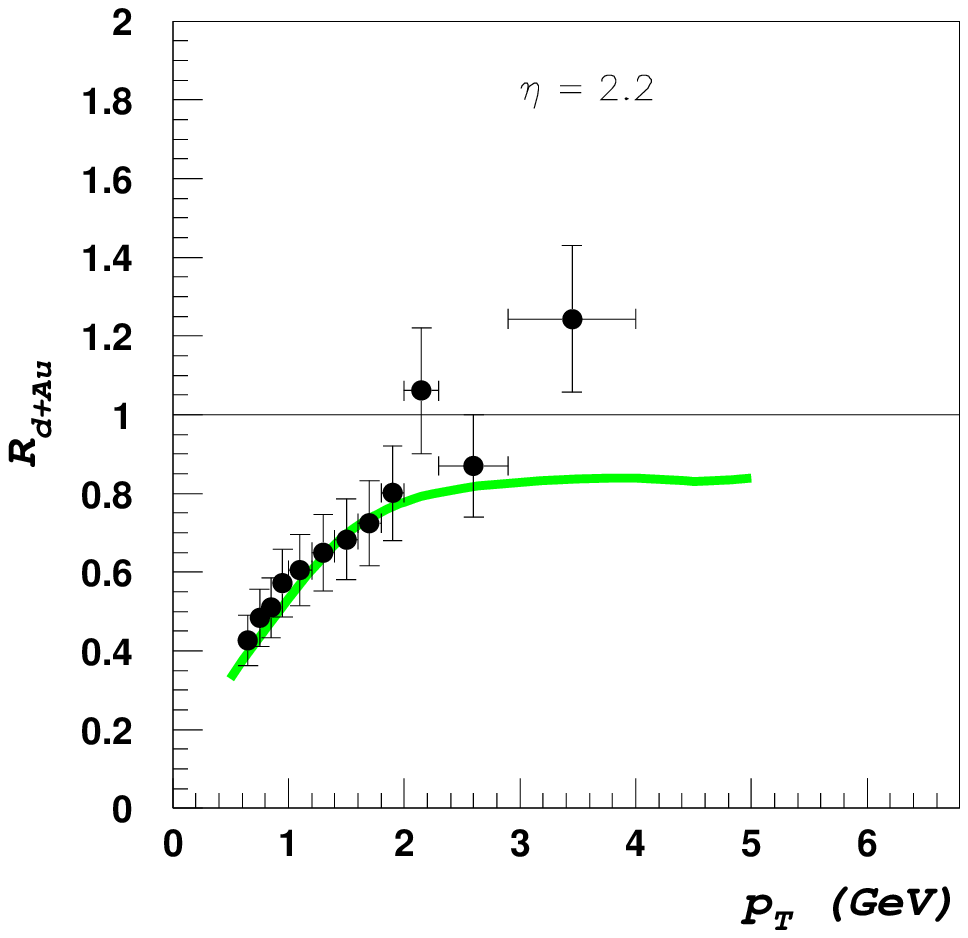}
\end{minipage}\hspace{1cm}
\begin{minipage}[b]{.35\linewidth}
\includegraphics[width=6cm]{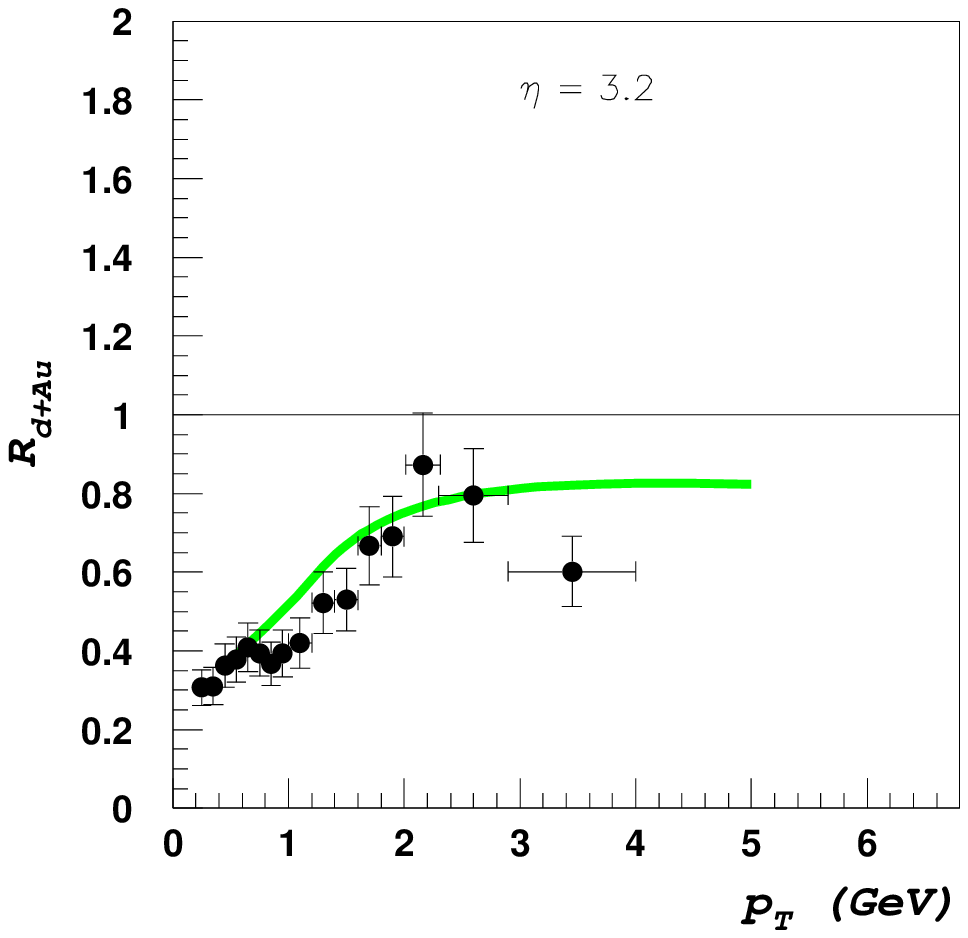}
\end{minipage}

\caption{
Nuclear modification factor in $dAu$ collisions as a function of transverse momentum for different rapidities; 
the data from BRAHMS Collaboration\protect\cite{Arsene:2004ux}, theoretical calculations from\protect\cite{Kharzeev:2003wz}.}
\vspace{-0.2cm}
\label{fig:brahms}
\end{wrapfigure}
Small $x$ evolution introduces anomalous dimension $\gamma \simeq 1/2$ in the gluon densities, so that the dependence on the momentum 
scale $Q$ is modified, to $Q^2 \to Q^{2 \gamma}$. Since in the vicinity of the saturation boundary the only dimensionful scale 
characterizing the system is the saturation momentum $Q_s^2$, the cross section of semi-hard scattering should scale as a function of 
$Q_s^2/Q^2$ -- it was found that this "geometrical scaling" \cite{Stasto:2000er,Iancu:2002tr,Mueller:2002zm,Armesto:2004ud} is consistent with HERA data on deep-inelastic scattering. Combining these two observations with the $A$ dependence of the saturation momentum $Q_s^2 \sim A^{1/3}$ we 
come to the conclusion\cite{Kharzeev:2002pc,Kharzeev:2003wz} that at sufficiently small $x$ and moderate $k_T$ the nuclear dependence of hard processes in $AA$ collisions should change from $S_A Q_s^4 \sim N_{part}^{4/3}$ (where $S_A \sim N_{part}^{2/3}$ is the overlap area) to $S_A Q_s^{4 \gamma} \sim N_{part}$. In $pA$ (or $dA$) 
collisions the nuclear dependence is  then $S_A Q_s^{2 \gamma} \sim A^{5/6}$, so there has to be a suppression as well. This suppression has 
also been found\cite{Albacete:2003iq} in the numerical solution of the Balitsky-Kovchegov equation, as well as in\cite{Baier:2003hr};  
for recent work, see also \cite{Blaizot:2004wu,Jalilian-Marian:2004xm,Iancu:2004bx}.

The experimental test of these ideas has been performed shortly afterwards --   
it has been established (for a review, see \cite{d'Enterria:2004fm} in this volume) that at mid-rapidity $y=0$ there is no high $k_T$ suppression in 
 $dAu$ data; this means that the suppression observed in $AuAu$ collisions has to come from the final-state effects, which will be discussed 
 below. The data thus rule out the possibility\cite{Kharzeev:2002pc} that $x$ is small enough for quantum evolution to develop already at mid-rapidity at RHIC. 
 Nevertheless, the presented arguments should apply at sufficiently small $x$.  
This is why the data on high $k_T$ hadron production at forward rapidities giving access to much smaller values of $x$ were eagerly awaited. 
The results from the BRAHMS experiment\cite{Arsene:2004ux} demonstrated a strong suppression of high $k_T$ hadrons; moreover, 
the centrality dependence appeared consistent with the predicted $R_{dA} \sim N_{part}^{-1/2}$ scaling, in a dramatic contrast 
to the increasing Cronin enhancement observed at mid-rapidity. Complementary results have been reported in\cite{Liu:2004kh,Back:2004bq,Barnby:2004yc}. 
STAR Colaboration has also reported\cite{Ogawa} on an observation of a predicted\cite{Kharzeev:2004bw} nuclear--dependent weakening  
of the back-to-back correlations for hadrons separated by several units of rapidity. It will be interesting to check if the suppression extends to charm hadrons at forward rapidities\cite{Kharzeev:2003sk}; at mid-rapidity, the first results have been reported in\cite{Tai:2004bf}. Alternative explanations based on "conventional" shadowing and multiple scattering (for a review, see\cite{Gyulassy:2003mc}) are also being explored.

\section{Approach to thermalization, and the r\^{o}le of classical fields}

There is by now an ample evidence of the importance of final state interactions in heavy ion collisions. 
Among the bulk observables, the azimuthal anisotropy of hadron production is a most spectacular 
evidence of this -- indeed, if all of the elementary nucleon--nucleon collisions were independent, 
the produced hadrons would not be correlated with the nucleus--nucleus reaction plane. 
The observed azimuthal anisotropy (or the "elliptic flow", in the parlance of the field; see \cite{Snellings,Voloshin:2002wa} for a review) 
indicates the existence of  
a correlation between the geometry of the nucleus--nucleus collision and the momenta of the emitted hadrons. 
An economical way of describing the evolution of a large number of particles in space and momentum is 
provided by relativistic hydrodynamics, which transforms the gradients of the initial parton density into 
the momentum flow of the produced hadrons. Hydrodynamical description is valid when the mean free path 
of partons is much smaller than the size of the system, i.e. when the system is sufficiently thermalized. 
The free expansion (or "inflation") of the produced system with time reduces the density gradients, 
so the magnitude of the elliptic flow crucially depends on the thermalization time when 
a hydrodynamical calculation is initiated. 
\begin{figure}
\noindent
\vspace{-0.3cm}
\includegraphics[width=16cm]{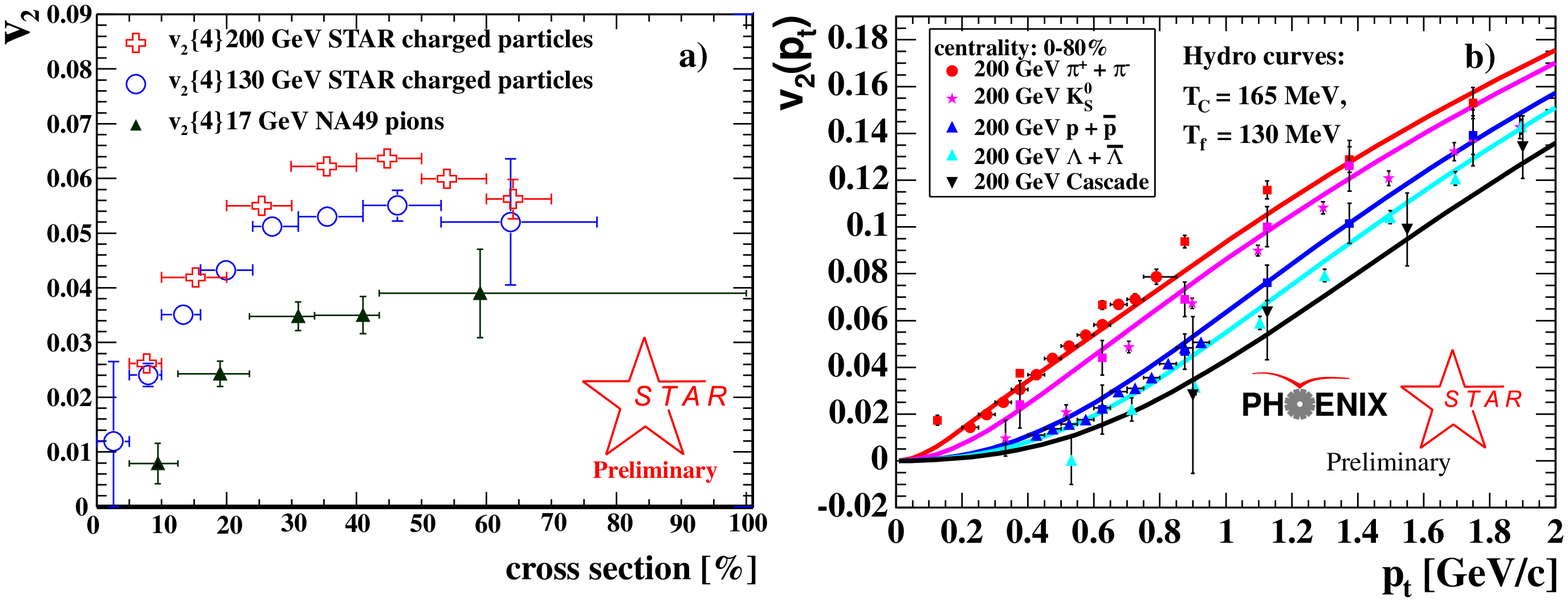}
\caption{
Elliptic flow near  mid-rapidity in Au + Au collisions as a function of centrality (left) and transverse momentum (right); from 
Ref.\protect\cite{Snellings}. }
\vspace{-0.5cm}
\label{fig:flow}
\end{figure}
It appears that to describe the elliptic flow of the observed magnitude \cite{Snellings},  one has to assume that 
the thermalization time is very short, about  $\tau_{therm} \simeq 0.5$ fm (for a review, see \cite{Heinz:2004pj,Shuryak:2004cy}). Such a short thermalization time presents 
a problem both for the traditional perturbative and non-perturbative treatments. Indeed, in perturbative QCD the 
rescattering amplitudes are suppressed by powers of the coupling $\alpha_s$, so the thermalization time appears 
long, on the order of $10$ fm, which makes the description of the elliptic flow problematic \cite{Molnar:2001ux}. 
In non-perturbative approaches, the interactions can be assumed strong, but the typical 
time scale of an interaction is $\sim 1/\Lambda_{\rm QCD} \sim 1$ fm, so it is difficult to expect that several interactions 
needed for thermalization will occur during  $\tau_{therm} \simeq 0.5$ fm.  The coherent classical fields present in the 
Color Glass Condensate (CGC) scenario may eventually provide a solution to this puzzle, since in this case the multi-gluon scattering  
amplitudes $A(n \to m)$ from $n \sim 1/\alpha_s$ to $m \sim  1/\alpha_s$ gluons are not suppressed. An approach to thermalization 
in this scenario was explored in Ref. \cite{Baier:2000sb}; recently, an attention was brought also to the role of instabilities in the 
equilibration process \cite{Arnold:2003rq}. The use of CGC initial conditions of Ref. \cite{Kharzeev:2000ph} in a hydrodynamical approach \cite{Hirano:2004rs}
has led to a successful description of the RHIC data. Nevertheless, much work will have to be done to understand
the thermalization process.

\section{Hydrodynamical evolution: more fluid than water}

Since hydrodynamical description relies on the direct use of the equation of state, the data can be used to extract an information 
on the properties of the medium. It appears that the quark--gluon plasma equation of state as measured on the lattice (for a review, see \cite{Karsch:2003jg}) is successful in describing the data. However the data can tell  even more about the properties of the 
medium, if one considers the influence of viscous corrections on various observables \cite{Teaney:2003pb}. Viscosity of the medium appears to affect the observables 
in a very significant way; in fact, one can deduce an upper limit \cite{Teaney:2003pb,Shuryak:2004cy} on the ratio 
of shear viscosity $\eta$ to the entropy density $s$, $\eta / s \leq 0.1$ -- much smaller than 
the same ratio for the water! Such a small value of viscosity, 
which reflects the dissipation of energy in a hydrodynamical evolution, contradicts  
the picture of weakly coupled quark-gluon plasma, and is more indicative of a 
strongly coupled quark-gluon liquid.
A calculation of shear viscosity in the strong coupling regime of QCD is still beyond the reach; however it has been made in $N=4$ supersymmetric Yang-Mills theory 
 \cite{Policastro:2001yc} -- the result is a small ratio of $\eta/s = 1/4 \pi$, comparable to the one inferred from RHIC data. 

A small value of viscosity in the strongly coupled quark-gluon plasma {\it a posteriori} justifies the use of the approach \cite{Kharzeev:2000ph} to hadron multiplicities assuming the proportionality of the number of measured hadrons to the 
number of the initially produced partons. This assumption would be unnatural if the evolution of the plasma were 
accompanied by parton multiplication, but 
is justified if the viscosity is small and evolution of the system is close to  
isenthropic.

\section{High $p_T$ hadron suppression, jet quenching, and heavy quarks}

The suppression of high $p_T$ hadrons in $Au-Au$ collisions is certainly one of the most spectacular new 
results at RHIC (for a comprehensive review of the data, see \cite{d'Enterria:2004fm} in this volume).  Such an effect has not been seen at lower energies\footnote{A moderate amount of suppression 
in the SPS results however cannot be excluded due to    
uncertainties in the reference $pp$ data  \cite{d'Enterria:2004ig}}; moreover, the results from the $d Au$ run at 
RHIC indicate that at pseudo-rapidity $\eta =0$ the observed suppression is entirely due to the final state effects, 
very likely a jet quenching in the quark-gluon plasma  
(for an overview, see  \cite{Gyulassy:2004vg,Jacobs:2004qv}). Alternative scenarios, e.g. the absorption  
in a dense hadron gas, seem unlikely in view of the high energy density $ \epsilon \sim 20\ {\rm GeV/fm^3}$ (see e.g. \cite{Kharzeev:2000ph}) achieved 
in the collisions. Nevertheless, additional experimental checks have to be performed; an important additional 
test of the jet quenching scenario involves the measurement of the suppression for heavy hadrons containing 
$c$ or $b$ quarks. If the suppression of high $p_t$ particles is indeed due to the induced radiation of gluons by 
fast partons, heavy quarks should lose significantly less energy than the light ones due to the "dead cone" 
effect \cite{Dokshitzer:2001zm}. This prediction seems to be in accord with the first RHIC data, 
which within the error bars indicate no quenching effect on the spectra of open charm, as inferred from the decay 
electrons \cite{Adcox:2002cg}; however more precise data are desirable. 
Several other calculations of the energy loss of heavy partons have been performed (see Refs.\cite{Djordjevic:2003qk,Armesto:2003jh,Thomas:2004ie} and papers \cite{Armesto:2004hz,Dainese:2004rh} in this volume); 
while they differ in the formalisms used, they all find a reduced energy loss for the heavy quarks. On the other hand, 
since heavy mesons ($D,B,...$) have a typical large size determined by the presence of the light quark in their 
 wave functions, in the hadronic absorption mechanism one would expect that heavy mesons interact with about the same 
 probability as the light ones.  

\begin{figure}[htb]
\noindent
\vspace{-0.8cm}
\begin{minipage}[b]{.46\linewidth}
\includegraphics[width=7.5cm]{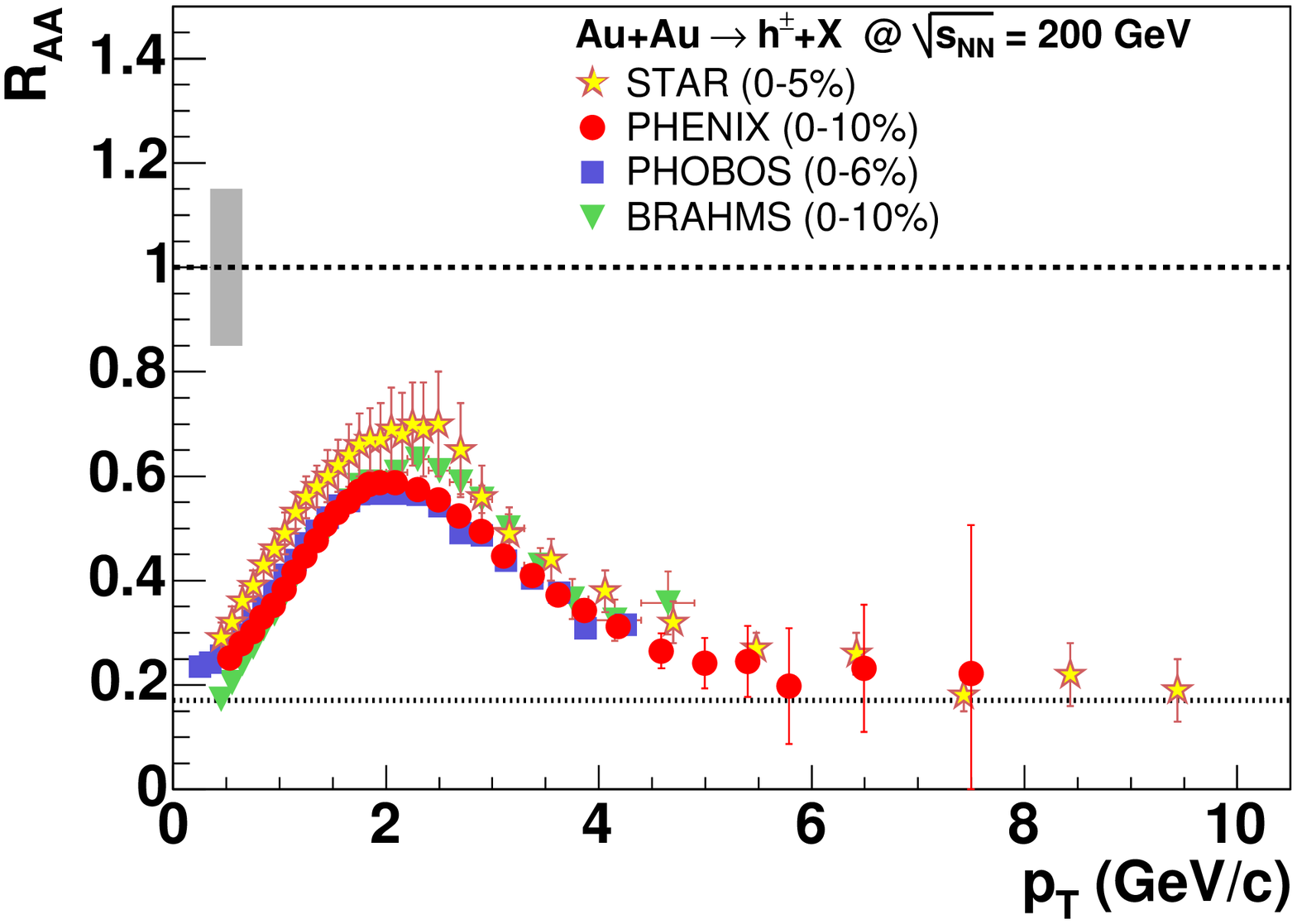}
\end{minipage}\hfill
\begin{minipage}[b]{.46\linewidth}
\includegraphics[width=7.5cm]{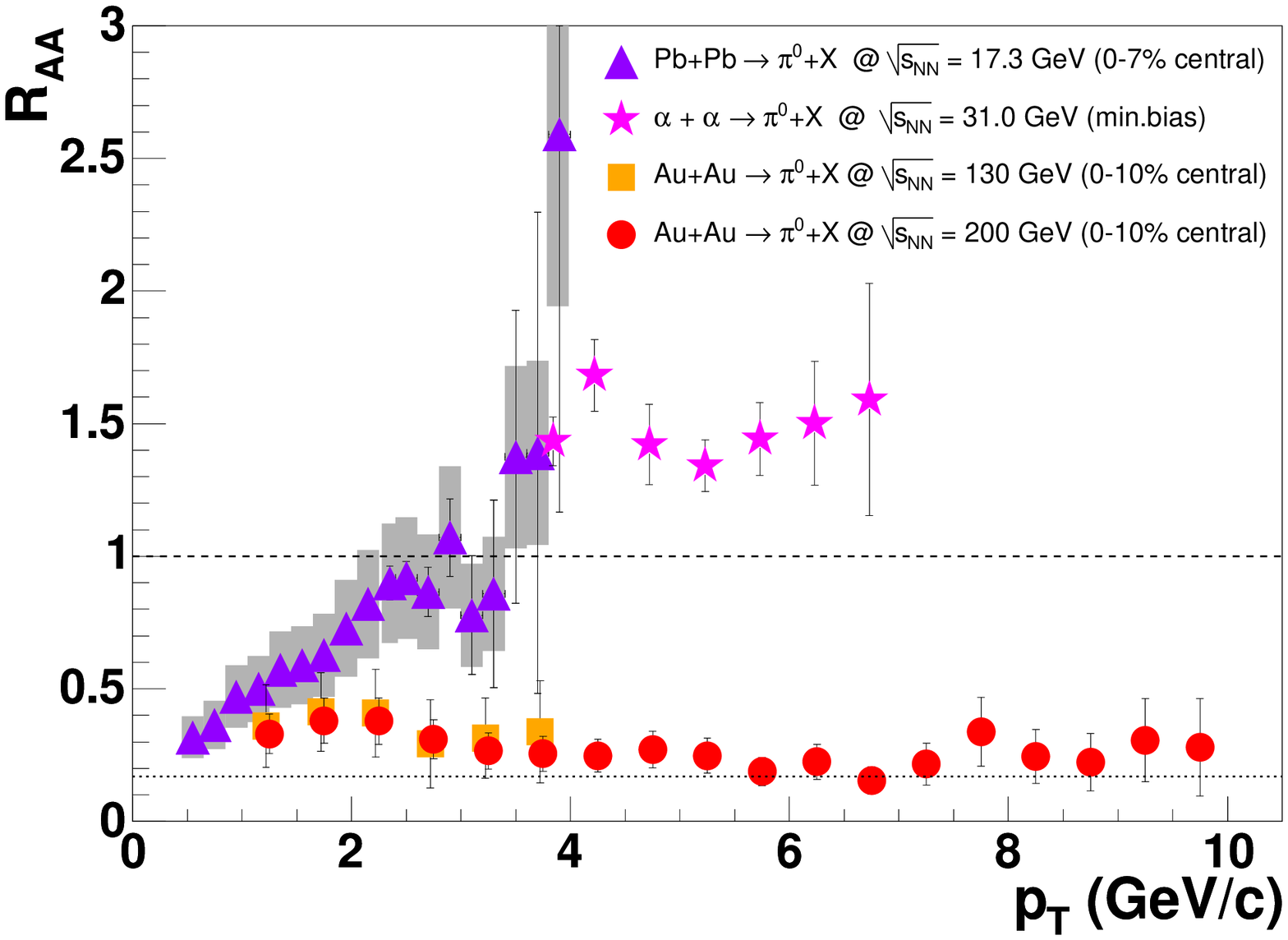}
\end{minipage}
\vspace{0.4cm}
\caption{
Nuclear modification factor for charged hadrons and neutral pions (left and right panels, respectively)
 in Au + Au collisions at $\sqrt{s} = 200$ GeV; from Ref.{\protect\cite{d'Enterria:2004fm}}}
\vspace{-0.5cm}
\label{fig:quench}
\end{figure}

\section{Heavy quarkonium in hot QCD matter}

Ever since it was proposed as a signature of the quark-gluon plasma\cite{Matsui:1986dk}, the dissociation of heavy quarkonia 
in hot QCD matter has remained a focal point of vigorous theoretical and experimental studies. 
The NA38/50 Collaborations 
at CERN have observed the suppression of $J/\psi$ and $\psi'$, and the current NA60 experiment will significantly extend the existing 
measurements (for an update on the recent results, see\cite{Arnaldi:2004wt} in this volume). The first RHIC results have already been reported,  in $AuAu$\cite{Adler:2003rc}, $dAu$\cite{deCassagnac:2004kb}, and $pp$\cite{Adler:2003qs} collisions. Theoretically, a new insight on the 
problem has been gained from the recent lattice calculations (see\cite{Datta:2004im,Asakawa:2003xj} for an overview) which indicate 
that the $J/\psi$ and $\eta_c$ survive as bound states in the quark-gluon plasma at least up to the temperatures twice the critical, $2 T_c$.  
This observation is very important in understanding the properties of the strongly coupled quark-gluon plasma discussed above. 
However, in my opinion, it should not be interpreted as an indication that heavy quarkonia are not suppressed in the quark-gluon 
plasma unless the temperature is very high; the point is that even if a $\bar{c}c$ state is bound in a plasma, it can be readily dissociated\cite{Shuryak:1978ij} 
by the impact of gluons, which have much harder momentum distributions in a deconfined phase\cite{Kharzeev:1994pz}. Estimates 
of the activation rate of quarkonia due to the interaction with the heat bath\cite{Kharzeev:1995ju} show that even if $(\bar{c}c)$ states exist as 
bound states, their yield can be strongly suppressed. More work has to be done to understand these effects better; on the lattice, a
reliable extraction of the thermal widths of heavy quarkonia would be most desirable.  

\section{Baryon dynamics}

The structure of baryons in non-perturbative QCD remains quite puzzling: while in non-relativistic quark model 
$QQQ$ baryons are not so different from $\bar{Q}Q$ mesons, in the approaches motivated by $1/N_c$ expansion 
they are drastically different -- in Skyrmion picture, for example, they are the topological solitons of the meson fields. 
A closer look at the quark wave functions of baryons reveals that local gauge invariance requires the presence of novel configurations of gauge field -- so called "baryon junctions"\cite{Rossi:1977cy}.  
Naively, one expects that at high energies the collision of two relativistic nuclei would not lead to any substantial baryon stopping -- 
since the valence quarks associated with the baryon number carry a large fraction of the nucleons' momentum, they are hard to stop in 
a soft process.  However the account of non-perturbative baryon junctions leads to a substantial change in this picture, since the baryon number 
appears to be traced by soft gluons\cite{Kharzeev:1996sq,Vance:1998vh,Arakelian:2002iw}. In perturbation theory, baryon junctions were 
shown to correspond to multi-gluon exchanges in higher color representations\cite{Kopeliovich:1988qm}. Substantial amount of baryon stopping, 
with the magnitude and rapidity dependence consistent with the baryon junction picture, has been observed at RHIC\cite{Bearden:2003hx}. 
The influence of quantum evolution and parton saturation on the $x$ distributions of valence quarks in nuclei has been addressed 
recently in Ref \cite{Itakura:2003jp}.

Another exciting observation at RHIC related to baryon dynamics is a strong enhancement of baryon-to-pion ratios at moderate values of 
transeverse momentum \cite{d'Enterria:2004fm} (so called $B/\pi$ puzzle) and the larger magnitude of the elliptic flow for baryons \cite{Snellings}. The proposed explanations 
include the phenomenon of parton coalescence\cite{Fries:2003kq,Molnar:2003ff,Hwa:2004mj} and the interplay of baryon junctions with 
jet quenching\cite{Vitev:2001zn}.

\section{Summary} 

The first years of the experiments at RHIC have changed in a dramatic way the theoretical picture of dense and hot parton systems. 
The evidence for the existence of new states of QCD matter is mounting\cite{Gyulassy:2004zy}, 
and a consistent description of the observed phenomena has started to emerge.   
Nevertheless, hot and dense QCD is still in its infancy -- and we have every reason to expect new surprises!  

\vskip0.1cm
I am grateful to the Organizers for the excellent meeting. 
This work was supported by the U.S. Department of Energy under Contract No. DE-AC02-98CH10886.

\end{document}